\documentclass[preprint,number]{elsarticle}
\usepackage{amssymb,amsmath}
\usepackage{hyperref}
\usepackage[usenames]{color}
\usepackage{bbold}
\usepackage{algorithm}
\usepackage{algpseudocode}
\usepackage{listings}
\DeclareMathOperator{\mathsp}{span}

\newcommand{\Kcal}{\mathcal{K}}

\begin{document}
\begin{frontmatter}
\title{Accelerating the solution of families of shifted linear systems with
CUDA}
\author{Richard Galvez}
\ead{ragalvez@syr.edu}
\address{Department of Physics, Syracuse University, Syracuse, NY 13244, USA}
\author{Greg van Anders}
\ead{grva@umich.edu}
\address{Michigan Center for Theoretical Physics, Randall Laboratory of Physics,
\\ The University of Michigan, Ann Arbor, MI 48109-1040, USA}
\begin{abstract}
  We describe the GPU implementation of shifted or multimass iterative solvers
  for sparse linear systems of the sort encountered in lattice gauge theory.
  We provide a generic tool that can be used by those without GPU programming
  experience to accelerate the simulation of a wide array of theories. We stress
  genericity, which is important to allow the simulation of candidate theories
  for new physics at LHC, and for the study of various supersymmetric theories.
  We find significant speed ups, which we conservatively bound below at at least
  twelve times, that promise to put a variety of research questions within
  practical reach.
\end{abstract}
\begin{keyword}
  CUDA \sep GPU \sep GPGPU \sep Krylov Subspace Methods
  \sep Lattice Gauge Theory
\end{keyword}

\end{frontmatter}
\section{Introduction} \label{sec:intro}
The solution of families of shifted linear systems is a problem that occurs in
many areas of scientific computing including partial differential equations
\cite{gallopoulossaad}, control theory \cite{dattasaad}, and quantum field
theory \cite{rhmc}. In the latter case, this problem is of particular importance
in the simulation of the strong coupling dynamics of gauge theories with
non-Abelian symmetry groups. Considerable computational effort has been devoted
to this problem, with particular emphasis on lattice studies of quantum
chromodynamics (QCD), the theory that describes the forces that bind atomic
nuclei together. In all such studies, the inclusion of dynamical fermions
results in a drastic increase in the cost of computation, but is key to
achieving physically meaningful results.

The rational hybrid Monte Carlo (RHMC) algorithm \cite{rhmc} is the most
efficient way of treating dynamical fermions in computer simulation. This method
approximates the Pfaffian or determinant, depending on the theory in question,
which arises through formal path integration over the fermionic fields, with a
related quadratic bosonic path integral that is diagonalized by solving a set of
shifted linear systems. Hence the main computational effort in simulating
strongly coupled non-abelian gauge theories is in solving a family of sparse
linear systems.

The most common technique for solving shifted families of sparse linear systems
is to use Krylov subspace methods (see section \ref{sec:krylov}). The fact that
Krylov subspace techniques can be used to simultaneously solve shifted families
of linear equations is a result of the simple fact that the Krylov subspace of a
matrix is not altered by shifting the matrix by some multiple of the identity.
It has been shown that a clever reuse of the iteration constants can generate a
whole family of shifted solutions without the need for additional matrix-vector
multiplications \cite{jegerlehner}. All solutions can therefore be generated at
the cost of generating the solution for which the linear problem is least well
conditioned.

Simulations, however, remain expensive and it is of great importance to find
methods for delivering results more quickly and cheaply. The advent of general
purpose computing using graphics processing units has led to the disruptive
acceleration of many areas of scientific computing. Examples abound, from
molecular dynamics \cite{hoomd} to atmospheric physics \cite{airpolutgpu}, in
which the use of GPUs to accelerate computation has allowed commodity desktop
hardware to rival the performance of conventional clusters with many CPU cores.
It seems is useful, then, to exploit this technology for the solution of
shifted linear systems.

In the case of simulations of QCD, considerable effort has been devoted to
developing techniques for delivering high performance simulations, including
methods for simulation using graphics processing units (GPUs)
\cite{lqcdvg,quda}. However, the problem of the simultaneous solution of shifted
linear systems is the key step in the simulation of a whole class of quantum
field theories, of which QCD is but one important member. Moreover, for the
study of theories other than QCD using code developed specifically for that
theory is prohibatively difficult. This represents an important problem, now
that high energy physics has entered the LHC era, there is the possibility that
interesting strongly coupled physics could turn out to be present at the TeV
scale. Were this to be the case, it would be important to have code that would
allow one to simulate candidate theories cheaply. Also, the simulation of
supersymmetric field theories is of great interest (see the review
\cite{Phys_Rep_Exact_SUSY}), not only because of the possibility that
supersymmetry could be revealed as a symmetry of nature. Supersymmetric theories
are of intrinsic interest themselves because of their rich mathematical
structure. They are also important for the testing and understanding of the
conjectured duality between certain gauge theories and gravity theories
motivated by considerations from string theory \cite{adscft1}. Reducing the
costs of both sorts of simulations requires the development of flexible tools
that can handle the bulk of the computational expense, and that can be easily
incorporated into the simulation of new theories as they become of interest.
Though existing GPU codes for QCD achieve remarkable performance, they are
designed specifically for studying that theory. It is therefore highly desirable
to have code that allows the solution of generic families of shifted linear
systems. Moreover, since this problem also occurs in partial differential
equations \cite{gallopoulossaad}, and in control theory \cite{dattasaad}, the
existence of stand-alone methods of sufficient genericity may be of use in those
fields as well.

In this paper we will describe a freely available open source code that can be
used to solve shifted families of sparse linear systems using NVIDIA GPUs.
Though this code is written in CUDA, its use does not require users to write
CUDA code directly. Instead they may simply call the routines without the
necessity of understanding how the code will be run on the GPU. The code
provides GPU implementations of the two most commonly used algorithms in
lattice gauge theory for simulating dynamical fermions: the multimass conjugate
gradient (CG-M) and multimass biconjugate gradient stabilized (BiCGStab-M)
methods. We find that our GPU solver can deliver the solution of a family of
linear systems with 15 shifts in less than 1/12$^\text{th}$ the time it takes to
deliver the solution the same system with no shifts on a single CPU core, using
our test hardware.

In section \ref{sec:krylov} we discuss the mathematical problem we wish to
solve, situating it in the context of simulations of quantum field theories
with dynamical fermions. We also review the key aspects of the iterative methods
used to solve problems of this class. In section \ref{sec:gpu} we describe the
the computational considerations that arise in solving problems in this class,
and describe how our implementation of the solution algorithm on the GPU handles
them. In section \ref{sec:perf} we describe the performance our programs. In
section \ref{sec:disc} we conclude with discussions for directions for future
work.

\section{Krylov Subspace Methods}\label{sec:krylov}
In the simulation of quantum field theories with fermions the bulk of the
computational expense involves performing the path integral over the fermions.
In a wide class of theories of interest, operators that occur in the Lagrangian
of the theory are quadratic in the fermions, meaning that they can be formally
integrated out, at the expense of leaving a purely bosonic path integral that
includes the Pfaffian or determinant of some operator whose precise form is
differs by theory. The calculation of the Pfaffian/determinant normally proceeds
by using a chain of identities in linear algebra and bosonic path integration to
covert the calculation into a quadratic path integral over a set of bosonic
fields, conventionally termed pseudofermions. The quadratic bosonic action is
difficult to write in a form that is suitable for doing simulations, i.e.\ in
diagonal form, because it involves the operator that appeared in the original
fermionic action, but now raised to some inverse fractional power. To
diagonalize the operator that appears in the quadratic pseudofermion action, the
standard practice is to approximate it by means of a series of rational
functions of the fermionic operator. This approximate form can be diagonalized
directly by solving a set of shifted linear systems.

At its essence the main computational hurdle in the simulation of quantum field
theories with dynamical fermions can be reduced to solving a set of linear
equations
\begin{equation}\label{eq:linprob}
  (A+\sigma_i I)x_i = b \, .
\end{equation}
In this expression $A$ is a matrix that is related to fermion operator in the
field theory, $I$ is the identity matrix, $\sigma_i$ are constants that come
from the rational approximation of the pseudofermion operator, $b$ is the
pseudofermion field, and $x_i$ are sought-after unknowns that contribute to
diagonalized form of the pseudofermion operator. In general, this system of
equations can get quite large. At the absolute lowest end, a recent study of
a one dimensional supersymmetric theory required the solution of complex linear
systems that were $1920^2$ \cite{pwmmsim}. Most applications, however, require
the solution of much larger systems. Reasonable simulations of maximally
supersymmetric Yang-Mills theory in four spacetime dimensions, for example,
require linear systems with sizes on the order $400000^2$ to study the theory
with gauge group $SU(2)$ \cite{a4st}. Moreover, this system must be solved
repeatedly throughout the simulation as new Monte Carlo moves are generated.
Fortunately, in most applications the matrix $A$ is sparse, which means that
iterative methods can be used, greatly reducing the computational expense.

There is a vast literature on iterative methods for the solution of sparse
linear systems; a nice review with references to key papers is given in
\cite{vandervorst}. These methods are of pivotal importance in, for instance,
the solution of discrete approximations of partial differential equations. A
common feature of many of them is that the solution makes use of the Krylov
subspace of the matrix $A$. Given some matrix $A$, and a vector $b$ (so chosen
because they will be precisely the quantities that appear in our problem
\eqref{eq:linprob}) the Krylov subspace of order $m$ of the matrix $A$ is given
by
\begin{equation}\label{eq:krylov}
  \Kcal_m(A) = \mathsp(A^k b,\forall k<m) \, .
\end{equation}
The iterative solution of linear problems proceeds by getting better
approximations to the exact solution by a carefully chosing the approximate
solution from successively higher order Krylov subspaces. In non-pathological
systems this converges to a unique solution. The $n^\text{th}$ order solution
can, in general, be written as
\begin{equation}\label{eq:poly}
  x^{(n)} = P_n(A) b
\end{equation}
where $P_n$ is some polynomial in the matrix $A$ whose coefficients are
different for different methods. A key feature of these methods is that the
most computationally expensive step is matrix-vector multiplication, which is
significantly less expensive than directly inverting the matrix.

For our purposes, we are interested in solving a family of related linear
problems that are all related by constant shifts. {\it A priori} it might seem
that we would be forced to solve each problem independently in turn. However,
one can notice from \eqref{eq:krylov} that
\begin{equation}
  \Kcal_m(A+\sigma_i I) = \Kcal_m(A) \, .
\end{equation}
This equivalence of the Krylov subspaces for the shifted and unshifted systems
suggests that it might be possible to simultaneously generate the solution of
the whole family of shifted solutions, provided one can construct shifted
versions of the polynomials appearing in \eqref{eq:poly}. Indeed, this has been
done explicitly in \cite{jegerlehner} for common solution algorithms. The cost
of producing the whole family of shifted solutions is determined by the cost of
solving the system that is least well conditioned \cite{jegerlehner}. In typical
applications, the shifts $\sigma_i$ are positive, meaning the shifted systems
are more well-conditioned then the unshifted one, and therefore the whole family
can be solved for the price of the solution of the unshifted system. For
illustrative purposes we present the CG-M algorithm \ref{alg:cgm}.
\begin{algorithm}
\caption{CG-M Algorithm}\label{alg:cgm}
\begin{algorithmic}[1]
\Procedure{CG-M}{$A,b,\sigma_i,\epsilon,N$}
\State $\alpha\gets0$
\State $\alpha_\sigma\gets 0$
\State $\beta_{0}\gets 1$
\State $\zeta_{-1}\gets 1$
\State $\zeta_{0}\gets 1$
\State $\zeta_{-1\sigma}\gets 1$
\State $\zeta_{0\sigma}\gets 1$
\State $r\gets b$
\State $p\gets b$
\State $p_\sigma\gets b$
\State $x_\sigma\gets 0$
\State $R_1\gets r^\dagger\cdot r$
\State $i\gets 0$
\While{$|r|>\epsilon$ and $i<N$}
  \State $R_0\gets R_1$
  \State $\beta_{-1}\gets\beta_0$
  \State $\bar p \gets A p$
  \State $P\gets p^\dagger \cdot \bar p$
  \State $\beta_0\gets R_0/P$
  \State $r\gets r+\beta_0 \bar p$
  \State $\zeta_{1\sigma}\gets\zeta_{0\sigma}\zeta_{-1\sigma}/
  (\zeta_{0\sigma}(1-\beta_0)+\zeta_{-1\sigma}(\beta_0-\sigma\alpha))$
  \State $\beta_{\sigma}\gets
  (\beta-\sigma\alpha)\zeta_{1\sigma}/\zeta_{0\sigma}$
  \State $R_1\gets r^\dagger \cdot r$
  \State $\alpha\gets R_1/R_0$
  \State $\alpha_\sigma\gets \alpha \zeta_{1\sigma}/\zeta_{0\sigma}$
  \State $p\gets r+\alpha p$
  \State $x_\sigma\gets x_\sigma-\beta_\sigma p_\sigma$
  \State $p_\sigma\gets \zeta_{0\sigma}r+\alpha_{\sigma}p_\sigma$
  \State $\zeta_{-1\sigma}\gets \zeta_{0\sigma}$
  \State $\zeta_{0\sigma}\gets \zeta_{1\sigma}$
  \State $i\gets i+1$
\EndWhile
\EndProcedure
\end{algorithmic}
\end{algorithm}
In this algorithm, variables indexed by $\sigma$ are arrays of values for each
of the desired shifts in the system of interest. Lines 2 through 14 initialize
the system. The variable $r$ is a vector that stores the residual for the
unshifted system defined by
\begin{equation}
  r = b-A\tilde x
\end{equation}
in which $\tilde x$ is the iterative solution of the unshifted system.
\footnote{Note that, as we are only interested in the solution of the shifted
systems, we needn't actually compute or store $\tilde x$ directly.} The criteria
for halting the iteration in line 15, are that the residual satisfies some
error tolerance, and that some iteration limit is not exceeded. The iterative
procedure is given in lines 16 through 32. A number of steps simply involve
reshuffling constants between iterations. Lines 22, 23 and 26 are responsible
for computing the iteration parameters for the shifted system in terms of those
of the unshifted system. The most expensive computational step is in line
18, where a matrix-vector product is computed. Lines 28 and 29 are key steps in
which the set of solutions of the shifted system are computed.

\section{Design of GPU Implementation}\label{sec:gpu}
For the iterative solution of linear systems the most expensive part of the
computation is sparse matrix-vector multiplication (SpMV). Achieving high
performance SpMV on the GPU was the focus of \cite{bellgarland} and the authors
of that paper have produced open source code implementing the ideas developed
therein \cite{cusp}. Although it was found that the SpMV routines did not
saturate the computing bound of the GPU, the authors were still able to see
drastically reduced wall-times for their code compared to a CPU implementation.

The aim of this paper is to develop routines to solve the family of problems
\eqref{eq:linprob} efficiently on the GPU. The key considerations affecting the
design of our implementation are: the code should be able to be called simply by
users who wish to incorporate the solver into existing code, or to incorporate
it into code used for studying new theories; the code should be written in a
programming language that would allow users to call it from a variety of
applications; users should have the option of having minimal knowledge of how
the computation is performed on the GPU; users should have the option of
developing highly optimized versions for specific problems if desired. As a
result of these considerations we were led to develop our solver using the
CUSP libraries \cite{cusp}.

CUSP is the aforementioned set of SpMV routines developed by the authors of
\cite{bellgarland}. These routines are, in turn, largely based on Thrust
\cite{thrust} a template library, much like the C++ standard template library,
but one which uses the GPU. These libraries provide a means by which we could
write a linear solver without requiring users to write any CUDA code, let alone
device kernels, directly. Indeed, much of the complication that arises in
GPU programming, including the allocation of memory on the GPU, and distribution
of parallel threads is handled by the library. Moreover, CUSP contains basic
sparse matrix and array containers that are can be easily incorporated into
existing code and filled with necessary data. Finally, it also allows the
flexibility to define custom linear operator kernels if specific applications
warrant the investment of time to develop highly optimized code.

The development of the solver was substantially aided by the existence of
efficient SpMV routines provided by CUSP, which we used to perform the
matrix-vector multiplication in line 18 of the algorithm \ref{alg:cgm}. The key
remaining performance consideration was to construct the rest of the solver in
such a way as to take advantage of the existing SpMV performance. The two main
issues requiring attention in this regard are the distribution of computation
between the CPU and GPU, and the construction of appropriate kernels to perform
computation on the GPU.

Regarding the distribution of computation betwen CPU and GPU, at various points
in the solution algorithm presented in section \ref{sec:krylov} part of the
iteration requires the computation of arrays of iteration constants, e.g.\ in
lines 22 and 23 of the algorithm \ref{alg:cgm}. The size of these arrays is
determined by the number of shifts, which for a typical application is on the
order of 10. Because the GPU exposes such a high degree of parallelism, such
computations represent a serial bottleneck. This does not, however, suggest that
one should those computations on the CPU. In general, copying memory between the
CPU and the GPU constitutes a substantial overhead, and it is preferable to
perform what is, in essence, a serial computation on the GPU.

We are led, therefore to implement the entire solution on the GPU. To do so
requires the construction of custom kernels to implement the various
calculations required during the iteration, of particular concern are the
kernels to implement the operations in lines 28 and 29 of the algorithm
\ref{alg:cgm}, which are the next most computationally expensive after the
SpMV operation. Some of these kernels are essentially the vectorization of level
one BLAS type operations, but others are more complicated. Given the expectation
that the routine would be memory-bandwidth limited, a key design consideration
for the kernels was the use of registers to store the data contained in array
elements that are used in multiple floating point operations.

We have implemented CG-M and BiCGStab-M solvers satisfying all of these
design criteria, which have been incorporated into the open-source CUSP project,
and are currently available online \cite{cusp}. An example program is shown
in figure \ref{fig:code}. In this code, line 3 includes the library containing
the solver. Lines 6 through 16 set the quantities that define the linear system.
In this case, line 15 loads a pre-defined matrix that is part of the CUSP
library. Line 18 defines a matrix in device memory, and copies the corresponding
matrix in host memory to it. Line 20 is the call to the solver. Line 22
allocates memory for the solution on the host, and copies the solution on the
device to it. Notice that this code does not involve explicit calls to CUDA
functions, rather they are hidden in the libraries that define the constructs
{\ttfamily cusp::device\_memory}.
\begin{figure}
\begin{center}
\begin{lstlisting}[language=C++]
#include <cusp/hyb_matrix.h>
#include <cusp/gallery/poisson.h>
#include <cusp/krylov/cg_m.h>
int main(void)
{
    typedef int    IndexType;
    typedef double ValueType;
    const size_t N = 1000;
    const size_t N_s = 15;

    cusp::hyb_matrix<IndexType,ValueType,cusp::host_memory> A_h;
    cusp::array1d<ValueType,cusp::device_memory> x_d(N*N_s,0);
    cusp::array1d<ValueType,cusp::device_memory> b(N,1);
    cusp::array1d<ValueType,cusp::device_memory> sigma(N_s,1);
    cusp::gallery::poisson5pt(A_h, N, N);
    cusp::default_monitor<ValueType> monitor(b, 2000, 1e-5);

    cusp::hyb_matrix<IndexType,ValueType,cusp::device_memory> A_d = A_h;

    cusp::krylov::cg_m(A_d, x_d, b, sigma, monitor);

    cusp::array1d<ValueType,cusp::host_memory> x_h = x_d;

    return 0;
}
\end{lstlisting}
\end{center}
\caption{An example call to the shifted conjugate gradient solver on the GPU
developed in this paper. Lines 18 and 22 copy data to and from the device, and
line 20 calls the solver.}
\label{fig:code}
\end{figure}

\section{Performance}\label{sec:perf}
To test the performance of our code we used a GPU equipped machine at the
Fermi National Accelerator laboratory (Fermilab). This machine has an Intel
Nehalem processor clocked at 2.67 GHz and 12 Gb of RAM, and an NVIDIA Tesla
S1070 card with 4 T10 GPUs. As a simple test of our method, we solved the
Poisson equation in two-dimesions simultaneously for a set of 15 shifts on the
GPU, and compared the peformance to the solution of a single unshifted system
on a single CPU core. We believe this provides a very conservative estimate of
the performance of our routine. The single unshifted solver we used for
comparison was the one from the CUSP library, and was written by others. This
ensures that any performance gains we report do not stem from us poorly
implementing the unshifted solver.

A test problem that has the virtue both of simplicity, and of demonstrating that
our routines are in no way specific to applications in high energy theory, is to
solve the Poisson equation in two-dimensions. We solve the 2d Poisson equation
by using a finite difference scheme in which the Laplacian is represented with
a five-point stencil on a $1000\times 1000$ grid. This system size is also
near the system sizes of interest for applications in high energy theory.

In table \ref{tab:stencilperf} we summarize the performance of our GPU
implementation of the CG-M algortithm compared to GPU and single-core CPU
implementations of the unshifted CG algorithm. We see considerable peformance
improvements of the shifted solver on the GPU compared with the unshifted solver
on a single CPU core. In double precision we see that our shifted solver can
produce the solution for a family of 15 shifts on the GPU in less than
$1/12^\text{th}$ the time it takes to produce the solution for a single shift
on a CPU core. Therefore, even by this conservative measure our solver is
performing well. Note however that the unshifted solver on the GPU is faster
than our shifted solver by a factor of a few. This is not unexpected and occurs
also for implementations on the CPU. In typical applications there are at least
a few shifts that do not lead to a significantly better conditioned system, and
so it remains computationally more feasible to use the CG-M routine once, rather
than the CG routine repeatedly.
\begin{table}\label{tab:stencilperf}
  \begin{center}
  \begin{tabular}{lll}
    \hline
    Routine & Single & Double \\
    \hline
    unshifted CG (Single Core CPU) & 
      & 159.158 ms/iteration \\
    unshifted CG (GPU) &
      2.20814 ms/iteration & 3.48276 ms/iteration \\
    CG-M with 15 shifts (GPU) &
      8.82461 ms/iteration & 12.9839 ms/iteration \\
    \hline
  \end{tabular}
  \end{center}
  \caption{Performance of unshifted CG solver on a single CPU core, and on the
  GPU \cite{bellgarland}, compared with the CG-M routine developed in this
  paper.}
\end{table}

\section{Discussion}\label{sec:disc}
We have presented results of an implementation of shifted Krylov subspace
solvers on the GPU. We demonstrated significant performance improvements over
similar solvers on the CPU. With an eye to future work in high energy physics,
the design of our solver stressed ease of incorporation into routines that would
allow the simulation of many different quantum field theories with minimal
modifications to existing code. We believe that our solvers constitute a
valuable tool as physicists contemplate the possibility of interesting strongly
coupled phenomena at the Large Hadron Collider, or attempt to recover black
hole thermodynamics from various strongly coupled supersymmetric gauge theories,
along the lines set out in \cite{cw}. Our shifted solvers CG-M and BiCGStab-M
are available through the open-source CUSP project. 

As we mentioned in the introduction, shifted linear systems are not only of
interest to quantum field theorists. Indeed, though they crop up in other areas
of scientific computing, they are a special case of a broader class of problems.One can imagine situations in which it would be desirable to solve a class of
linear problems given by
\begin{equation}\label{eq:linprobgen}
  (A+\sigma_i I)x_i = b_i \, .
\end{equation}
I.e.\ those similar to \eqref{eq:linprob}, but with a different right-hand side
for each shift. {\it A priori}, this would appear problematic, since the
usual initial choice of solution would put $x_i=0$, and therefore the residuals
$b_i$ would, in general be linearly independent. This would mean that the
resulting Krylov subspaces would differ between shifts. It was pointed out in
\cite{osborn}, however, that a judicious choice of initial guesses can deliver
the same initial residual for each of the linear systems. This implies, in turn,
that the same Krylov subspace method could be used to solve this family of
systems as well. The method for the generation of initial guesses
in \cite{osborn} involves a number of SpMV operations, and so it seems
worthwhile to develop a routine capable of producing these intial guesses using
the GPU, after which our solver could finish the solution of the family of
systems.

\section*{Acknowledgements}
We would like to thank S.~Catterall and S.~Glotzer for helpful discussions,
and N.~Bell, J.~Hoberock, and D.~Holmgren for helpful communication, 
J.~ Anderson for comments on the manuscript, and M.~van Anders for
collaboration at the beginning of this work. This work is supported in part
by the US Department of Energy under grant DE-FG02-95ER40899. Tests were
performed using USQCD resources at Fermilab. 


\begin{thebibliography}{10}
\expandafter\ifx\csname url\endcsname\relax
  \def\url#1{\texttt{#1}}\fi
\expandafter\ifx\csname urlprefix\endcsname\relax\def\urlprefix{URL }\fi
\expandafter\ifx\csname href\endcsname\relax
  \def\href#1#2{#2} \def\path#1{#1}\fi

\bibitem{gallopoulossaad}
E.~Gallopoulos, Y.~Saad, Efficient parallel solution of parabolic equations:
  Implicit methods on the cedar multicluster, in: J.~Dongarra, P.~Messina,
  D.~C. Sorensen, R.~G. Voigt (Eds.), PPSC, SIAM, 1989, pp. 251--256.

\bibitem{dattasaad}
B.~N. Datta, Y.~Saad, Arnoldi methods for large sylvester-like observer matrix
  equations, and an associated algorithm for partial spectrum assignment,
  Linear Algebra and its Applications 154-156 (1991) 225 -- 244.
\newblock \href {http://dx.doi.org/10.1016/0024-3795(91)90378-A}
  {\path{doi:10.1016/0024-3795(91)90378-A}}.

\bibitem{rhmc}
M.~A. Clark, A.~D. Kennedy, Z.~Sroczynski, Exact 2+1 flavour rhmc simulations,
  Nucl. Phys. Proc. Suppl. 140 (2005) 835--837.
\newblock \href {http://arxiv.org/abs/hep-lat/0409133}
  {\path{arXiv:hep-lat/0409133}}.

\bibitem{jegerlehner}
B.~Jegerlehner, {Krylov space solvers for shifted linear systems}\href
  {http://arxiv.org/abs/hep-lat/9612014} {\path{arXiv:hep-lat/9612014}}.

\bibitem{hoomd}
J.~A. Anderson, C.~D. Lorenz, A.~Travesset, General purpose molecular dynamics
  simulations fully implemented on graphics processing units, Journal of
  Computational Physics 227~(10) (2008) 5342 -- 5359.
\newblock \href {http://dx.doi.org/10.1016/j.jcp.2008.01.047}
  {\path{doi:10.1016/j.jcp.2008.01.047}}.

\bibitem{airpolutgpu}
F.~Moln\'ar, T.~Szak\'aly, R.~M\'esz\'aros, I.~Lagzi, Air pollution modelling
  using a graphics processing unit with cuda, Computer Physics Communications
  181~(1) (2010) 105 -- 112.
\newblock \href {http://dx.doi.org/10.1016/j.cpc.2009.09.008}
  {\path{doi:10.1016/j.cpc.2009.09.008}}.

\bibitem{lqcdvg}
G.~I. Egri, Z.~Fodor, C.~Hoelbling, S.~D. Katz, D.~Nogradi, K.~K. Szabo,
  {Lattice QCD as a video game}, Comput. Phys. Commun. 177 (2007) 631--639.
\newblock \href {http://arxiv.org/abs/hep-lat/0611022}
  {\path{arXiv:hep-lat/0611022}}, \href
  {http://dx.doi.org/10.1016/j.cpc.2007.06.005}
  {\path{doi:10.1016/j.cpc.2007.06.005}}.

\bibitem{quda}
M.~A. Clark, R.~Babich, K.~Barros, R.~C. Brower, C.~Rebbi, {Solving Lattice QCD
  systems of equations using mixed precision solvers on GPUs}, Comput. Phys.
  Commun. 181 (2010) 1517--1528.
\newblock \href {http://arxiv.org/abs/0911.3191} {\path{arXiv:0911.3191}},
  \href {http://dx.doi.org/10.1016/j.cpc.2010.05.002}
  {\path{doi:10.1016/j.cpc.2010.05.002}}.

\bibitem{Phys_Rep_Exact_SUSY}
S.~Catterall, D.~B. Kaplan, M.~Unsal, {Exact lattice supersymmetry}, Phys.
  Rept. 484 (2009) 71--130.
\newblock \href {http://arxiv.org/abs/0903.4881} {\path{arXiv:0903.4881}},
  \href {http://dx.doi.org/10.1016/j.physrep.2009.09.001}
  {\path{doi:10.1016/j.physrep.2009.09.001}}.

\bibitem{adscft1}
J.~M. Maldacena, The large {N} limit of superconformal field theories and
  supergravity, Adv. Theor. Math. Phys. 2 (1998) 231--252.
\newblock \href {http://arxiv.org/abs/hep-th/9711200}
  {\path{arXiv:hep-th/9711200}}.

\bibitem{pwmmsim}
S.~Catterall, G.~van Anders, {First Results from Lattice Simulation of the
  PWMM}, JHEP 1009 (2010) 088.
\newblock \href {http://arxiv.org/abs/1003.4952} {\path{arXiv:1003.4952}},
  \href {http://dx.doi.org/10.1007/JHEP09(2010)088}
  {\path{doi:10.1007/JHEP09(2010)088}}.

\bibitem{a4st}
S.~Catterall, R.~Galvez, G.~van Anders, Work in progress.

\bibitem{vandervorst}
H.~A. van~der Vorst, Iterative Krylov methods for large linear systems,
  Cambridge University Press, Cambridge, UK, 2003.

\bibitem{bellgarland}
N.~Bell, M.~Garland, Implementing sparse matrix-vector multiplication on
  throughput-oriented processors, in: SC '09: Proceedings of the Conference on
  High Performance Computing Networking, Storage and Analysis, ACM, New York,
  NY, USA, 2009, pp. 1--11.
\newblock \href {http://dx.doi.org/10.1145/1654059.1654078}
  {\path{doi:10.1145/1654059.1654078}}.

\bibitem{cusp}
N.~Bell, M.~Garland, Cusp: Generic
  parallel algorithms for sparse matrix and graph computations, version 0.1.0
  (2010).
\newline\urlprefix\url{http://cusp-library.googlecode.com}

\bibitem{thrust}
J.~Hoberock, N.~Bell.
\newblock \href{http://www.meganewtons.com/}{Thrust: A parallel template
  library} [online] (2009).
\newblock Version 1.1.

\bibitem{cw}
S.~Catterall, T.~Wiseman, {Towards lattice simulation of the gauge theory duals
  to black holes and hot strings}, JHEP 12 (2007) 104.
\newblock \href {http://arxiv.org/abs/0706.3518} {\path{arXiv:0706.3518}},
  \href {http://dx.doi.org/10.1088/1126-6708/2007/12/104}
  {\path{doi:10.1088/1126-6708/2007/12/104}}.

\bibitem{osborn}
J.~C. Osborn, {Initial guesses for multi-shift solvers}, PoS LATTICE2008 (2008)
  029.
\newblock \href {http://arxiv.org/abs/0810.1081} {\path{arXiv:0810.1081}}.

\end{thebibliography}

\end{document}